\documentstyle[psfig,aps,amsmath,preprint,tighten]{revtex}

\title{Eigen's error threshold and mutational meltdown in a 
 quasispecies model}
\author{Franco Bagnoli\cite{fb}}
\address{Dipartimento di Matematica Applicata,
Universit\`a di Firenze \\
via S. Marta, 3 I-50139, Firenze, Italy.\\
Also INFN and INFM, sez. di Firenze.}
\author{Michele Bezzi\cite{mb}}
\address{Dipartimento di Fisica,
Universit\`a di Bologna \\
Via Irnerio, 46,
I-40126 Bologna, Italy.\\
Also INFN, sez. di Bologna}

\begin{document}

\maketitle
\begin{abstract}
 We introduce a toy model for interacting populations connected by
 mutations and limited by a shared resource. We
  study the presence of Eigen's error threshold and
 mutational meltdown. The phase diagram of the system shows that the
 extinction of the whole population 
 due to mutational meltdown can occur well before
 an eventual error threshold transition.
\end{abstract}
\vspace*{10pt}

{\bf keywords:}  Speciation models; 
 Darwinian Theory; Population 
Dynamics; Eigen Model; Mutational Meltdown.

The evolutionary process is due to the balance between two opposite forces:
random mutation
and natural selection. Mutations tend to increase the genetic diversity,
while selection
discriminates the individuals  more apt to survive.
 A simple model for evolution of
self-reproducing molecules was proposed by
Eigen~\cite{Eigen}. The inheritable characters (genotype)
are modeled by a binary string $g=(g_1, g_2,\dots,g_L)$, $g_i\in
\{0,1\}$, of fixed length $L$ (haploid individuals).
The genotypic space 
is thus a binary hypercube of $L$ dimensions. One can think of the zeros as
good genes (or basis) and the ones as bad genes.
We shall denote the set of individuals sharing the same genotype 
as a strain.

The selection is modeled by the concept of  fitness, which can be
defined as the survival probability or birth 
rate of a strain in the
limit of zero mutations and vanishing population (in order to avoid
overcrowding effects).   
The selection acts on the external characters of an individual, i.e.\
its {\em phenotype} $m$, which can be considered a (generally degenerated)
function of the genotype $m(g)$. The simplest phenotype (used also in Eigen's
model)  is a decreasing function of the fraction of bad genes in the
genotype: $m(g)=(1/L)\sum_i g_i$ . The mutations, occurring with
probability $\mu$, reverse the value of a single bit (point mutations) and 
thus couple different strains. 

The classification of strains into species is based both on the
phenotypic traits and on the genotypic information: for strains to
belong to the same species, they must be connected by mutations and
form an isolated cluster in the phenotypic space. This last
requirement is guaranteed, even in an almost flat fitness landscape, 
by competition~\cite{BagnoliBezzi}, which
``eliminates'' the less adapted strains phenotypically near to a fittest
one.  

For a  \emph{sharp peak landscape} the 
genotype $g=(0,0,\dots)$ 
has higher fitness $\alpha$, and all 
other genotypes have the same lower fitness $\beta$: the fitness landscape is
almost flat.  
Two different regimes are possible: in one phase
the asymptotic limit is a bell shaped distribution 
in phenotypic space (quasispecies),
centered around the master sequence; in the other
the most common phenotype is no more the fittest one. 
The shape of the distribution
in this second phase is dominated by combinatorial factors,
approaching a Gaussian in the limit of very large mutation rates, 
with almost no difference among strains. 
    
The transition between these two regimes, triggered by 
the mutation rate $\mu$ or
by the length of the genome $L$,
is called \emph{error threshold}. While the original results have been obtained
for an infinite population, this transition, 
disregarding the effects of fluctuations, depends slightly on the size of
population~\cite{NowakSchuster,Alves}. The error transition has the character of a true phase transition
(first order) for an infinite genome length $L$~\cite{Galluccio}. 

A related effect, effective in small populations, is called Muller's
ratchet~\cite{Maynard} or \textit{stochastic
escape}~\cite{Higgs95,Wood96}: in absence of back mutations there is a
finite probability that the master sequence will be lost due to
fluctuations. Since it relies on a random process, 
the average escape time is
however of order of the exponential of the size of the
strain~\cite{Pal}, 
and thus this effect is relevant only for very small
populations. 
In practice, this is the ultimate extinction mechanism,
effective when another cause has reduced enough the population.
The error threshold, on the other hand, does not depend on the
size of the total population. 
In the following we shall neglect to consider the influence of
fluctuations.

When the total population size is not kept fixed, the presence of unfavorable
mutations reduces the average fitness, and this can lead to
the extinction of the whole population, a phenomenon which 
 is called \emph{mutational meltdown}  
~\cite{Lynch90,Lynch93,Bernardes}. The presence of this effect in the Eigen
model has been shown numerically in Ref.~\cite{Malarz}.   

In this work we propose a simple minimal model which is able to 
exhibit both the error threshold and the mutational meltdown. 
Due to the form of the fitness function, the dynamics of the
population is fundamentally determined by the fittest strains. Let us indicate
with $X$ the number of individuals sharing the master sequence, 
 with $Y$ the
number of individuals whose genotype has $m=1$, and with $Z$ all
others individuals. 
We assume also non-overlapping generations, so that we can consider
a discrete time dynamics with unit equal to the generation time. 

During reproduction, individuals from strain $X$ can mutate, 
contributing to $Y$, and $Y$ to $Z$,   
with mutation rate $\mu L$. We disregard the possibility of back mutations from
$Z$ to $Y$ and from $Y$ to $X$. 
This last assumption is
equivalent to the limit of a large genome, which is the case for existing
organisms. We also introduce
the reproduction accuracy $q$, defined in term of $\mu L $ as $q=1-\mu
L$. Due to the assumption of large $L$, the multiplicity of
mutations from $m=1$ to $m>1$ ($L-1$)
is almost the same of that from $m=0 $
to $m-1$ ($L$). 

We shall assume a finite (and constant) 
carrying capacity $K$ of the environment, 
assuming that the effective reproduction 
rate of a population is proportional to
$1-N/K$, where $N=X+Y+Z$ is the total number of individuals. 
The evolution equation of the population is
\begin{equation}
\left\{
\begin{array}{rcl}
 X' &=&\left(1-\dfrac{N}{K}\right)q \alpha X, \\
 Y' &=&\left(1-\dfrac{N}{K}\right)\left(
  q \beta Y+(1-q) \alpha X\right),\\
 Z' &=&\left(1-\dfrac{N}{K}\right)\left(
  \beta Z+(1-q) \alpha Y\right);
\end{array}
\right.
\label{xy}
\end{equation}
and 
\[
	N'=\left(1-\dfrac{N}{K}\right)(\alpha X + \beta (Y+Z)),
\]
where the prime denote quantities at following time step.
Obviously all population dies if $\alpha < 1$. 

By introducing the normalized distribution $x=X/N$, $y=Y/N$  and $%
z=Z/N $ ($x+y+z=1$) we have
\begin{equation}
\left\{
\begin{array}{rcl}
 x' &=&\dfrac{q \alpha x}{\alpha x+\beta (y+z)},\\ 
 y' &=&\dfrac{q \beta y+(1-q) \alpha x}{\alpha x+ \beta (y+z)},\\
 z' &=&\dfrac{\beta z+(1-q) \alpha y}{\alpha x+ \beta (y+z)},
\end{array}
\right.
\label{evol}
\end{equation}
and
\begin{equation}
	N' = N\left(1-\dfrac{N}{K}\right)(\alpha x+\beta (1-x)).
\end{equation}

The steady state of Eq.~(\ref{evol}) is given by $x'=x$, 
$y' = y$ and 
and $z'=z$. We obtain two fixed points
\begin{equation}
	A = \left\{\begin{array}{rcl}
		x^{A} &=&0, \\
		y^{A} &=&0,\\
		z^{A} &=& 1;
	\end{array}\right.\label{A} 
\end{equation}
and
\begin{equation}
	B = \left\{\begin{array}{rcl}
		x^{B} &=&\dfrac{q \alpha -\beta}{\alpha -\beta}, \\
		y^{B} &=&\dfrac{(1-q)}{q}\dfrac{ \alpha(q\alpha
			-\beta)}{(\alpha -\beta)^2}, \\
		z^{B} &=&\dfrac{(1-q)^2}{q} \dfrac{\beta\alpha}{(\alpha
			-\beta)^2}. 
	\end{array}\right.
	\label{B} 
\end{equation}
The fixed point $A$ is always unstable and $B$ always stable for $%
\alpha >1$. In this second case the asymptotic population size is
\begin{equation}
	N^B = K\left(1-\dfrac{1}{q\alpha}\right).
	\label{NB}
\end{equation}

We characterize the error threshold 
 by the fact that  the fittest {\em phenotype} is no longer
the most common one, its phase boundary is given by  $x=y$. 
Evaluating this condition at the fixed point $B$ we get from
Eqs.~(\ref{B})
\begin{equation}
	q_{e} =\dfrac{\alpha}{2\alpha -\beta}. \label{qe}
\end{equation}

The mutational
meltdown  
corresponds to the vanishing of the total population $N=0$,
while not changing
its distribution. Thus, from Eq.~(\ref{NB}) we have
\begin{equation}
	q_{m} =\dfrac{1}{\alpha}. \label{qm}
\end{equation}

Our definition of the error threshold transition
needs some remarks: in the original work~\cite{Eigen} the error
threshold is
located at the maximum mean Hamming distance, which
corresponds to the maximum spread of population. In the limit of very
large genomes these two definitions agree, since the transition
becomes very sharp~\cite{Galluccio}. For finite populations, 
a possible transition can  be located
in correspondence of the vanishing of the
probability of finding the master
sequence~\cite{NowakSchuster,Malarz}, in a way similar to Muller's
ratchet. 

However, since the master sequence has always an higher effective
fitness than other strains, this implies that for $\beta < 1$
the error threshold is
only a transient phenomenon before extinction, unless 
the population is artificially kept 
finite, as in Ref.~\cite{NowakSchuster}. This mutation-induced
extinction has been investigated numerically by Malarz and Tiggemann 
in Ref.~\cite{Malarz}. 

The condition for the disappearance of the master sequence is $x'/x
< 1$ in the limit of vanishing $x$. Denoting arbitrarily this transition
as Muller's ratchet's one, we obtain from the first of
Eqs.~(\ref{evol})
\begin{equation}
	q_r = \frac{\beta}{\alpha}.\label{qr}
\end{equation}
This transition coincides with the mutational meltdown 
 for $\beta=1$.

In Figure~1 we plot the phase diagram of the three transitions
for some
different values of $\beta$. 
We note that the error threshold depends only on the ratio
$\alpha/\beta$ (i.e.\ on the relative fitness of strains);
on the other hand, the mutational
meltdown threshold involves the whole population, so it depends on the absolute
value of the fitness $\alpha$. 
For $\beta<1$ the mutational meltdown
transition crosses the error threshold, so one can  observe both
extinction of the quasispecies distribution or the error threshold in
a stable population,
according with the parameters. 

The physical relevant conditions
for the extinction of the
master sequence is $\beta<1$, for which also the $Y$ strains vanish;
otherwise back mutations cannot be neglected.  
Our toy model can approximate the M2 model by Malarz and Tiggemann,
in which only one bit is mutated with probability $P_{\mbox{mut}}$, and thus  
$q=1-P_{\mbox{mut}}$. In that model the selective advantage $\alpha/\beta$ 
of the
master sequence is given by the number of offsprings,
$N_{\mbox{rep}}$, while $\beta$ is given by the reproduction
probability $P_{\mbox{rep}}$. Since they observed the extinction of the master
sequence as error threshold, their data must be compared with
Eq.~(\ref{qr}), 
thus
\[
	P_{\mbox{mut}} = 1-\dfrac{1}{N_{\mbox{rep}}}
\]
corresponds to their error threshold transition and 
\[
	P_{\mbox{mut}} = 1-\dfrac{1}{P_{\mbox{rep}}N_{\mbox{rep}}}
\] 
to the mutational meltdown. The phase diagram of this model is
reported in Figure~2. The numerical data of Ref~\cite{Malarz} are well
reproduced, except for a small shift of the error threshold
transition.  

In summary we find that for a very steep fitness function ($\alpha \gg \beta$),
by increasing the mutation rate $\mu$ or the genome length $L$ 
one always observes the error threshold; but for
moderately fitness difference among strains 
the mutational meltdown causes the extinction of the
whole population while retaining the quasispecies distribution.

\begin{figure}
\centerline{
\psfig{figure=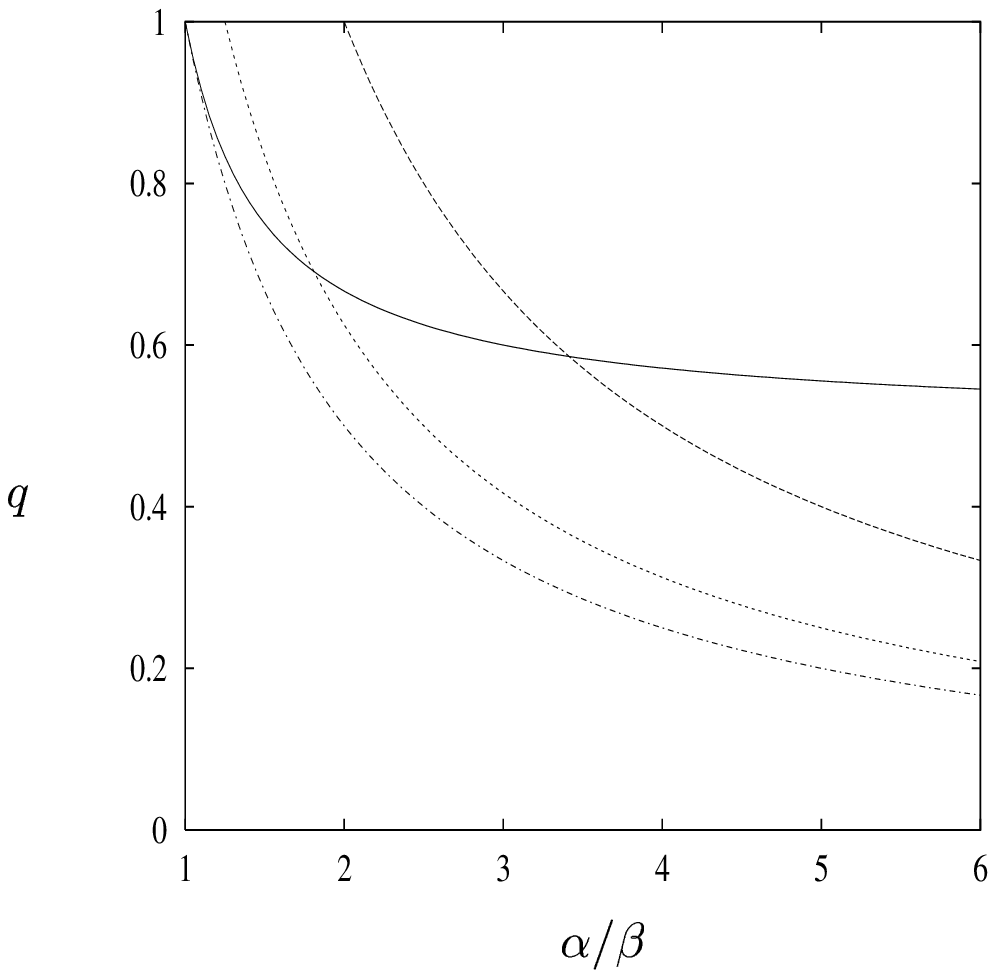,angle=0,width=16cm}
}
\caption{Phase diagram for the error threshold and mutational meltdown
transitions. The continuous line corresponds to the error threshold
$q_e$, Eq.~(\protect\ref{qe}), the dashed line to the mutational
meltdown $q_m$, Eq.~(\protect\ref{qm}), with $\beta=0.5$, the dotted
line to 
the mutational
meltdown with $\beta=0.5$, the dashed-dotted line to 
the mutational
meltdown with $\beta=1.0$, corresponding to Muller's ratchet
$q_r$, Eq.~(\protect\ref{qr}). 
Since $q=1-\mu L$, the increasing of 
the mutation rate or the genome
 length corresponds to lowering $q$. The lower left corner  
correspond to extinction.}
\label{toy:fig1}
\end{figure}

\begin{figure}
\centerline{
\psfig{figure=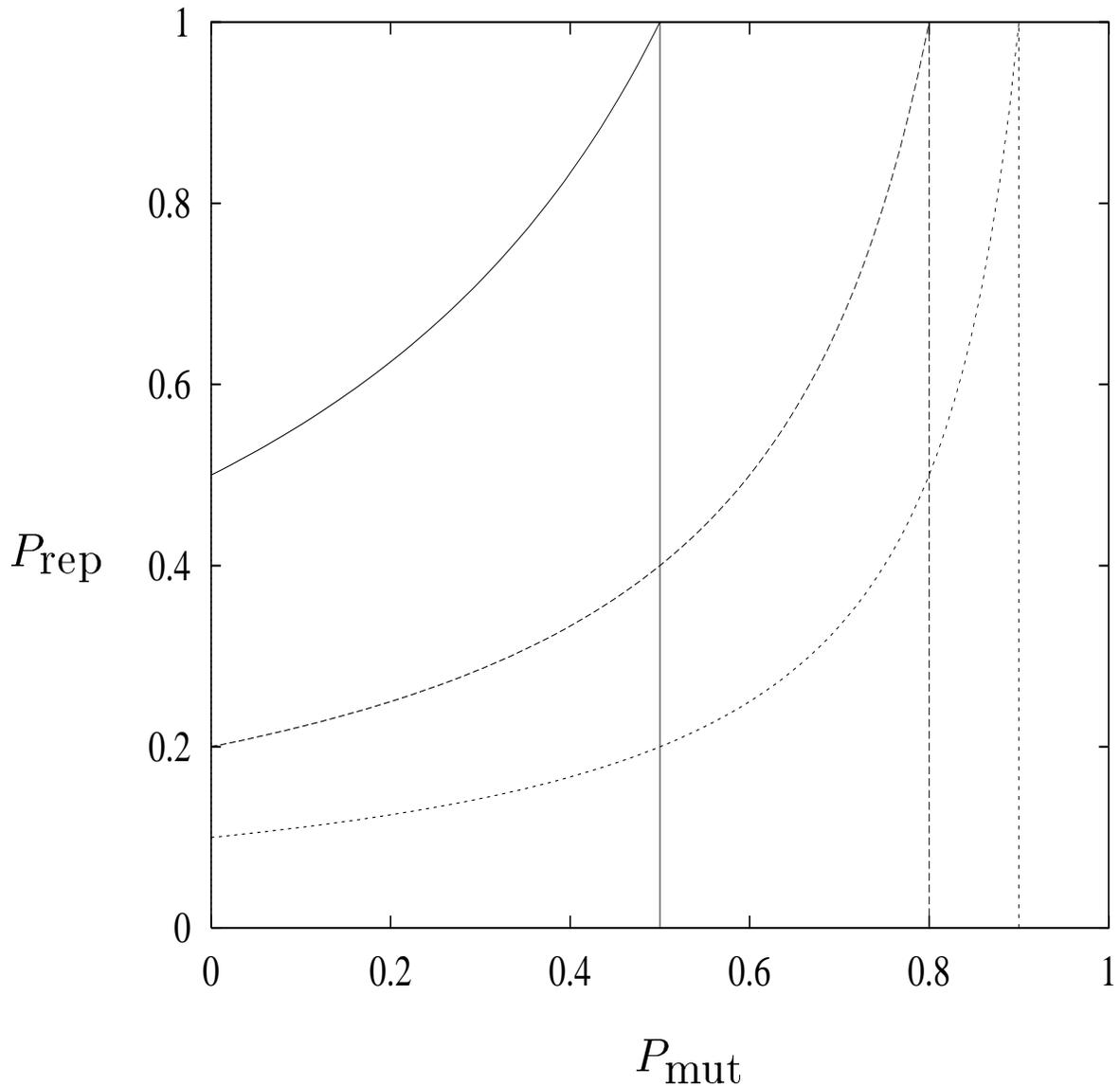,angle=0,width=16cm}
}
\caption{Phase diagram for the error threshold (Muller's ratchet
version, vertical lines) and mutational meltdown (curves) 
transitions for model M2 of Ref.~\protect\cite{Malarz}:
$\alpha = P_{\mbox{rep}}N_{\mbox{rep}}$, $\beta= P_{\mbox{rep}}$, 
$q= 1-P_{\mbox{mut}}$. Continuous line $N_{\mbox{rep}}=2$, dashed line 
$N_{\mbox{rep}}=5$, dotted line $N_{\mbox{rep}}=10$. 
}
\label{toy:fig2}
\end{figure}

\end{document}